\documentclass[aps,prl,amsmath,amssymb,floatfix,superscriptaddress,twocolumn]{revtex4}
\usepackage{subfigure}
\usepackage{amsmath}
\usepackage{graphicx}

\begin{document}

\title{Interference of nematic quantum critical quasiparticles: a route to the octet model}

\author{Eun-Ah Kim}
\affiliation{Department of Physics, Cornell University, Ithaca, NY 14853}
\author{Michael J. Lawler}
\affiliation{Department of Physics, Binghamton University, Binghamton NY 13902}
\affiliation{Department of Physics, Cornell University, Ithaca, NY 14853}

\date{\today}


\begin{abstract}
Repeated observations of inhomogeneity in cuperate superconductors\cite{MAR2000, Howald:2001vn, Lang:2002kx, Vershinin:2004rt, Kohsaka:2007fk} make one immediately question the existance of coherent quasiparticles(qp's)
and the applicability of a momentum space picture. Yet, obversations
of interference effects\cite{McElroy:2003mz, Kohsaka:2008fj, Doiron-Leyraud:2007ys,Sebastian:2008lr}
suggest that the qp's maintain a remarkable coherence under
special circumstances. In particular, quasi-particle interference
(QPI) imaging using scanning tunneling spectroscopy revealed a highly
unusual form of coherence: accumulation of coherence only at special
points in momentum space with a particular energy dispersion\cite{McElroy:2003mz,
Kohsaka:2007fk,Kohsaka:2008fj}. Here we show that nematic quantum critical fluctuations\cite{Kim:2008gf}, combined with the known extreme velocity anisotropy\cite{Chiao:2000ve}
provide a natural mechanism for the accumulation of coherence at those
special points. Our results raise the intriguing question of whether
the nematic fluctuations provide the unique mechanism for such a phenomenon.
\end{abstract}
\maketitle

The capability of QPI studies in inferring momentum space electronic structure
from real space local density of states(LDOS) images is surprising given the
strong presence of glassiness and nanoscale inhomogenaity in cuprate superconductors.
The simplicity of the QPI image, a set of well defined dispersing peaks, is particularly striking considering the complexity of the real space image. \textcite{McElroy:2003mz} made an insightful observation: the peak positions are determined by the eight tips of the ``banana'' shaped
qp equal energy contours. However, a key question remains of this ``octet model'' --- 
what makes the qp's at the tips especially coherent to
the extent that only interference among qp's at the
tips remain visible.

Underlying the QPI interpretation of the Fourier
transform of the LDOS map $N(\vec{r},\omega)$, is
the assumption that the modulated contribution $\tilde N_{\rm imp}(\vec{q},\omega)$ results from 
 coherent qp's scattering off sparcely
distributed impurities. For conventional metals where these assumptions
hold, the modulations in LDOS can be understood in terms of interference
among single particle wave functions\cite{Crommie:1993ul}. Naturally, there has been much effort towards 
interpreting the QPI in cuprates also in terms of free (Bogoliubov)
qp pictures \cite{Wang:2003zr, Pereg-Barnea:2003ly, Balatsky:2006yq}. 
Surprisingly, these free QPI patterns
are far more intricate than the set of simple dispersing peaks observed in the experiment. Hence one might
view this as a rare circumstance in which an experimental message
is much simpler than that provided by the simplest theory. 

\begin{figure*}
\includegraphics[width=0.8\textwidth]{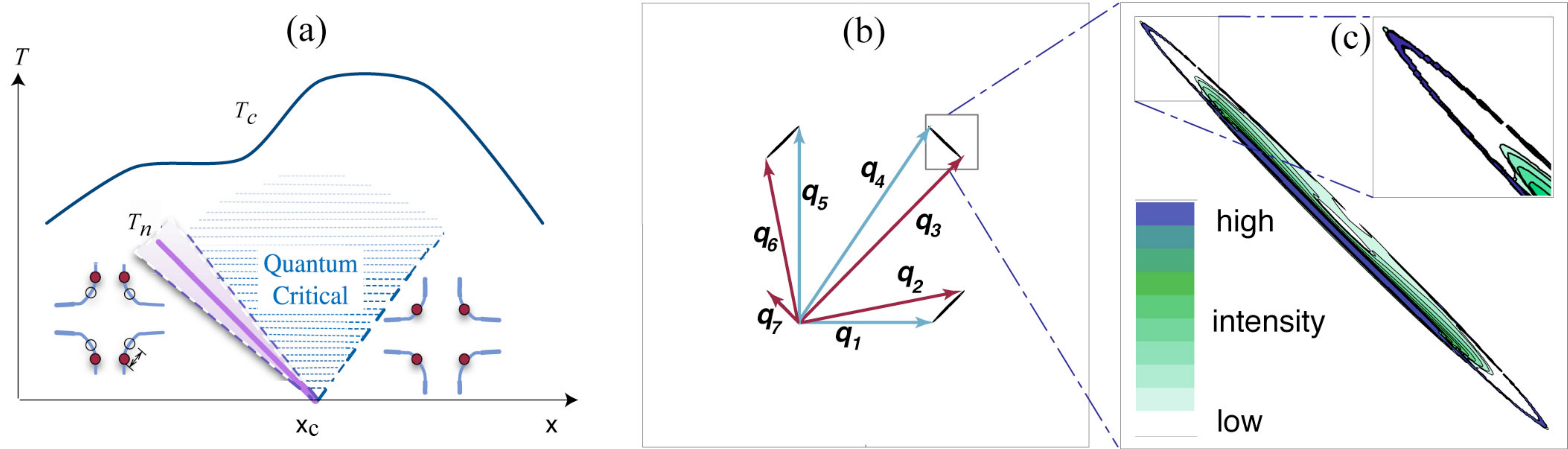}
\caption{
The electron spectral distribution at the nodal nematic quantum critical point in the linearized approximation (taken from Ref. \cite{Kim:2008gf}). (a) proposed phase diagram inside the superconducting dome, where $x$ is a tuning parameter. Note the change in location of the nodes through the phase transition. (b) Equal energy contour at $\omega=$3meV and the connecting $\vec q$-vectors of the banana tips. The $\vec{q}_i$ are colored blue and red depending on whether they connect $\vec{k}$ points with the same or opposite sign of the $d$-wave gap $\Delta_{\vec{k}}$. 
   (c) blowup of the momentum distribution $A(\vec k,\omega=\text{3meV})$ of the spectral function near one node. Note the sharpness of the momentum distribution at the banana tip shown in the inset.
}\label{fig:NN}
\end{figure*}

What is necessary to explain QPI in the cuprates, given these free
theory results, is to introduce scattering among the electrons in
such a way that only the banana tip qp's remain coherent.
Recently, toegether with our collaborators, we have shown\cite{Kim:2008gf}
that such a phenomenon naturally occurs at the nodal nematic
quantum critical point (QCP) of a d-wave superconductor. ``Nematic''
here refers to a broken symmetry phase in which the fourfold rotational
symmetry of the crystal is broken down to a twofold symmetry (see Fig. \ref{fig:NN}(a)). In a
d-wave superconductor, such additional symmetry breaking results in
a shifting of the nodal positions away from their fourfold symmetric
locations\cite{Vojta:2000fr}.
At the nodal nematic QCP, we found\cite{Kim:2008gf} that the softening of the nodal positions
introduces strongly $\vec{k}$ dependent decoherence 
that brings in stark contrast between the tips of the banana, where qp's remain coherent with long lifetime, and the rest of the equal energy contour, where qp's  get severely damped (see Fig. \ref{fig:NN}(b) and (c)).

Here we show that $\vec{k}$ dependent decoherence due to nematic fluctuations
leads to the very simplification
observed in QPI experiments:  peaks in fourier transform LDOS.
We further argue for the uniqueness of this route based on the severely restricted qp scattering mechanisms in a d-wave superconductor  due to their limited phase space\cite{Vojta:2000fr} combined with the following experimental evidence supporting the existence of nematic ordering in underdoped cuprates. \textcite{V.Hinkov02012008} found direct evidence for nematic ordering in 
YBa$_2$Cu$_3$O$_{6.45}$. Furthermore, both the glassy nematic  behavior in underdoped Bi$_2$Sr$_2$CaCu$_2$O$_{8+\delta}$\cite{Kohsaka:2007fk} and the doping dependent flattening of the near-node gap slope  Bi$_2$Sr$_2$CaCu$_2$O$_{8+\delta}$\cite{Tanaka:2006rg,Le-Tacon:2006le,Kohsaka:2008fj} can be naturally explained by  an increasing degree of nematicity upon underdoping.

In a many-body setting, a crisp way to capture the modulation of LDOS
resulting from isolated impurities is to use the T-matrix formalism (see for example Ref. \cite{Balatsky:2006yq})
which expresses the energy resolved LDOS $\tilde N_{imp}(\vec{q},\omega)$
at wave vector $\vec{q}$ as 
\begin{equation}
\tilde N_{imp}(\vec{q},\omega)=-2\text{sgn}(\omega)\text{Im}\int d\vec{k}\left[\hat{\mathcal{G}}(\vec{k}+\vec{q},\omega)\hat{T}\hat{\mathcal{G}}(\vec{k},\omega)\right]_{11}
\label{eq:Nimp}
\end{equation}
where $\hat{\mathcal{G}}(\vec{k},\omega)$ is the ($2\times2$)
single particle Nambu matrix propagator in a superconducting state
without impurity scattering and $\hat{T}$ is the impurity potential
in the weak (perturbative) impurity limit. Here $\hat T$ depends on the type of impurity. For a charge impurity which is antisymmetric in Nambu space, $\hat{T}=V_c\hat{\sigma}_3$, while $\hat{T}=V_m\mathbb{I}$ for a magnetic impurity and $\hat{T}=V_\Delta\hat{\sigma}_1$ for a superconducting gap impurity, both of which are symmetric in Nambu space ($\hat{\sigma}_i$ are Pauli matrices acting on Nambu spinor).

Equation\eqref{eq:Nimp}  shows  that
$\tilde N_{imp}(\vec{q},\omega)$ is the amplitude modulation associated
with the overlap between two qp states with wave vector $\vec{k}$
and $\vec{k}+\vec{q}$ scattering off the impurity. Hence $\tilde N_{imp}(\vec{q},\omega)$
will show high intensity at a special vector $\vec{q}$, if it connects
two coherent (long lived) qp states with the same energy
$\omega$ constructively. 
\textcite{McElroy:2003mz} and \textcite{Wang:2003zr}
noticed that the density of state $Im\hat{\mathcal{G}}$ is accumulated at the  $\vec{k}$
points at the eight tips of the ``banana'' contours. 
In fact,  a comparison between auto-correlation analysis of ARPES spectra and the QPI study of the same system showed remarkable similarity demonstrating this principle\cite{mcelroy:2006tq}. However, this similarity is not reproduced by calculations of relevant quantities.  Hence  peaks observed in QPI requires a mechanism that goes beyond accumulation of states and makes the tips especially coherent. Such mechanism has been a theoretical mystery. 

The nature of the propagating single qp states is encoded in the Nambu matrix propagator $\hat{\mathcal{G}}$ entering equation\eqref{eq:Nimp}. If, in between impurity scattering events, the qp's experience collisions with critical nematic collective modes, a self energy  $\hat{\Sigma}$ is induced (see e.g., Ref. [\onlinecite{Schrieffer:1999pd}])
\begin{equation}
\hat{\mathcal{G}}^{-1}=\hat{\mathcal{G}}_{0}^{-1}-\hat{\Sigma}.
\label{eq:G}
\end{equation}
Here $\hat{\mathcal{G}}_{0}$ is the free Bogoliubov qp propagator
of a BCS superconductor. In order to capture the nematic critical fluctuations, we use the self-energy obtained in 
Ref.\cite{Kim:2008gf}.

In the context of
the cuprates, $\hat{\mathcal{G}}_{0}(\vec{k},\omega)$ takes the form
\begin{equation}
\hat{\mathcal{G}}_{0}^{-1}=(\omega+i\delta)\mathbb{I}-\varepsilon_{\vec{k}}\sigma_{3}-\Delta_{\vec{k}}\sigma_{1}\label{eq:G0}
\end{equation}
where $\varepsilon_{\vec{k}}$ is the dispersion of normal state qp's  and 
 $\Delta_{\vec{k}}=\Delta_{0}(\cos k_{x}a-\cos k_{y}a)$ is 
 the d-wave pairing amplitude.  
In the low energy long wavelength limit, we can approximate this $\hat{\mathcal{G}}_{0}$
by linearizing around the four gapless nodal points $\vec{k}\approx\vec{K}$
where the qp energy $\xi_{\vec{k}}=\sqrt{\varepsilon_{\vec{k}}^{2}+\Delta_{\vec{k}}^{2}}$
vanishes:
\begin{equation}
\mathcal{G}_{0}^{-1}\bigg|_{\text{near node }\vec{K}}=(\omega+i\delta)\mathbb{I}-v_{F}\big(k_{x}-K_{x}\big)\hat{\sigma}_{3}-v_{\Delta}\big(k_{y}-K_{y}\big)\hat{\sigma}_{1}
\label{eq:G0-lin}.
\end{equation}
Linearzing  $\varepsilon_{\vec{k}}$ and $\Delta_{\vec{k}}$ given in Ref.[\onlinecite{Norman:1995bh}] based on photoemission data, we find
 $v_{F}=0.508$ and $v_{\Delta}=0.025$ in units of $eV(a/\pi)$.
(Note that the resulting anisotropy ratio $v_{F}/v_{\Delta}=20.3$ is large and consistent
with the value of 19 inferred from thermal conductivity measurements\cite{Chiao:2000ve}.)

\begin{figure*}
\subfigure[]{\includegraphics[width=0.3\textwidth]{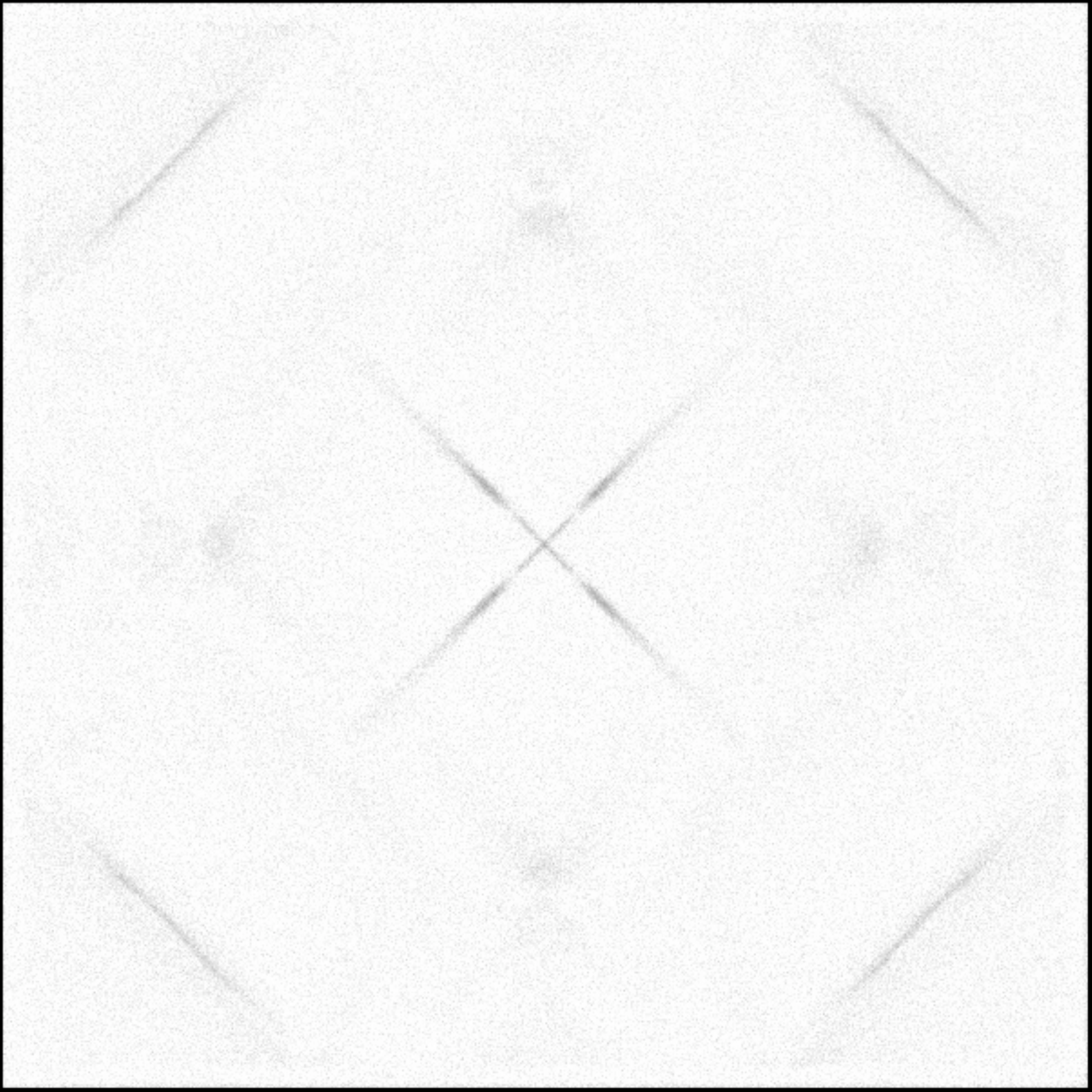}}
\subfigure[]{\includegraphics[width=0.3\textwidth]{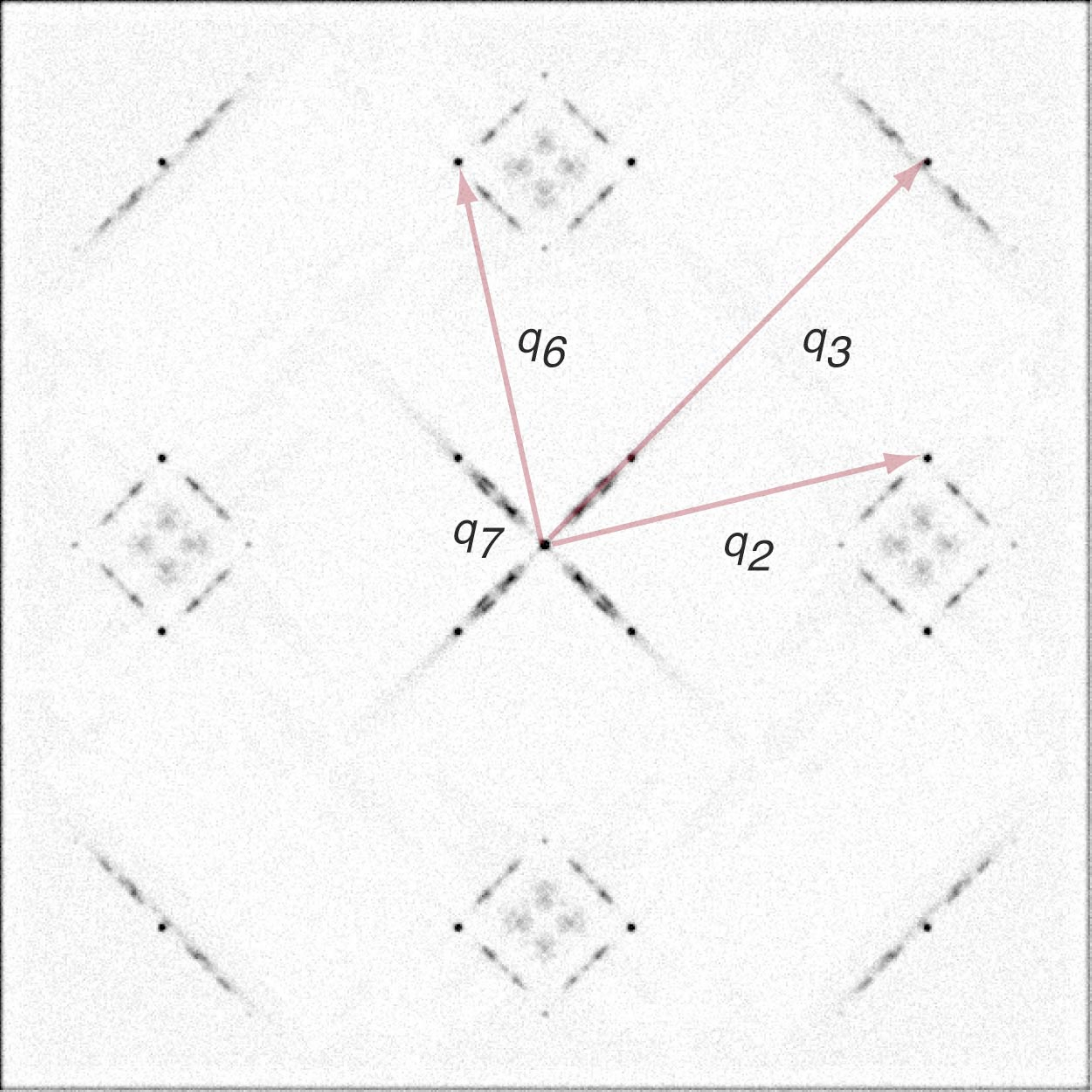}}
\subfigure[]{\includegraphics[width=0.3\textwidth]{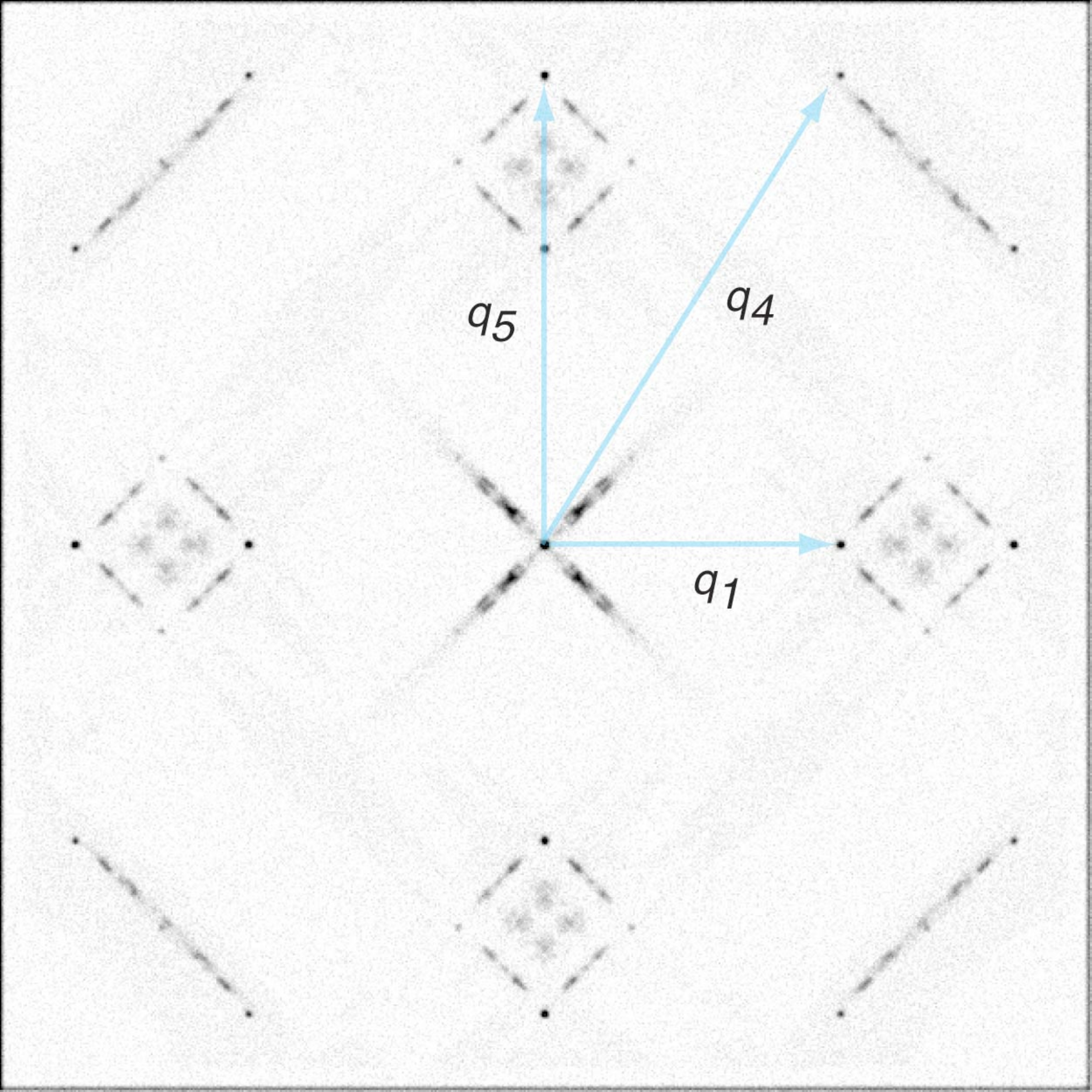}}
\caption{
(a) QPI Fourier amplitudes  $|N(q_{x},q_{y},\omega=9\text{meV})|$ for free Bogoliubov qp's
with linearized dispersion. (b) QPI Fourier amplitudes at the nematic QCP in the presence of scalar (non-magnetic) impurities ($\vec q_i$, $i=\{2,3,6,7\}$ label the constructively interfering peaks)
(c) QPI Fourier amplitudes at the nematic QCP in the presence of magnetic impurities ($\vec q_i$, $i=\{1,4,5\}$ label the constructively interfering peaks). The $\vec{q}_{i}$ vectors, defined in Fig. \ref{fig:NN}(b) are precisely those specified in the octet model of Ref. \onlinecite{McElroy:2003mz} . Note that at all $\vec q_i$ the QPI Fourier amplitudes are dramatically enhanced at the Nematic QCP. We used a gray scale where black represents 
 intensities greater than a fixed threshold that is the same for all plots. 
}\label{fig:qpi}
\end{figure*}

The effect of nematic critical fluctuations in the QPI intensity is evident when compared with
free linearized Bogoliubov qp's.
In Fig.\ref{fig:qpi}(a), we plot the QPI intensity $\tilde N_{imp}(\vec{q},\omega=9meV)$  for  the free qp's with $\hat{\mathcal G}=\hat{\mathcal G_0}$ induced by  charge impurities $\hat{T}=V_c\hat{\sigma}_3$. Most noticeable features in Fig.\ref{fig:qpi} (a) are the extended and broad
line shaped segments and the less extended but faint patterns. The vectors
$\vec{q}_{i}$ connecting the tips of bananas shown in Fig. \ref{fig:NN}(b) are
overlaid on the intensity plot in Fig.\ref{fig:qpi}(b) and (c). Clearly, $\vec{q}_{3}$, $\vec{q}_{4}$
and $\vec{q}_{7}$ land on the broad line shaped segments but the
rest of  $\vec{q}_{i}$ vectors
point to very faint features.
This is inconsistent with the experiments
showing well defined peaks at all $\vec{q}_{i}$ vectors albeit with varying intensities.

\begin{figure}[t]
\subfigure[Free Bogoliubov qp's, charge impurities
]{\includegraphics[width=0.49\columnwidth]{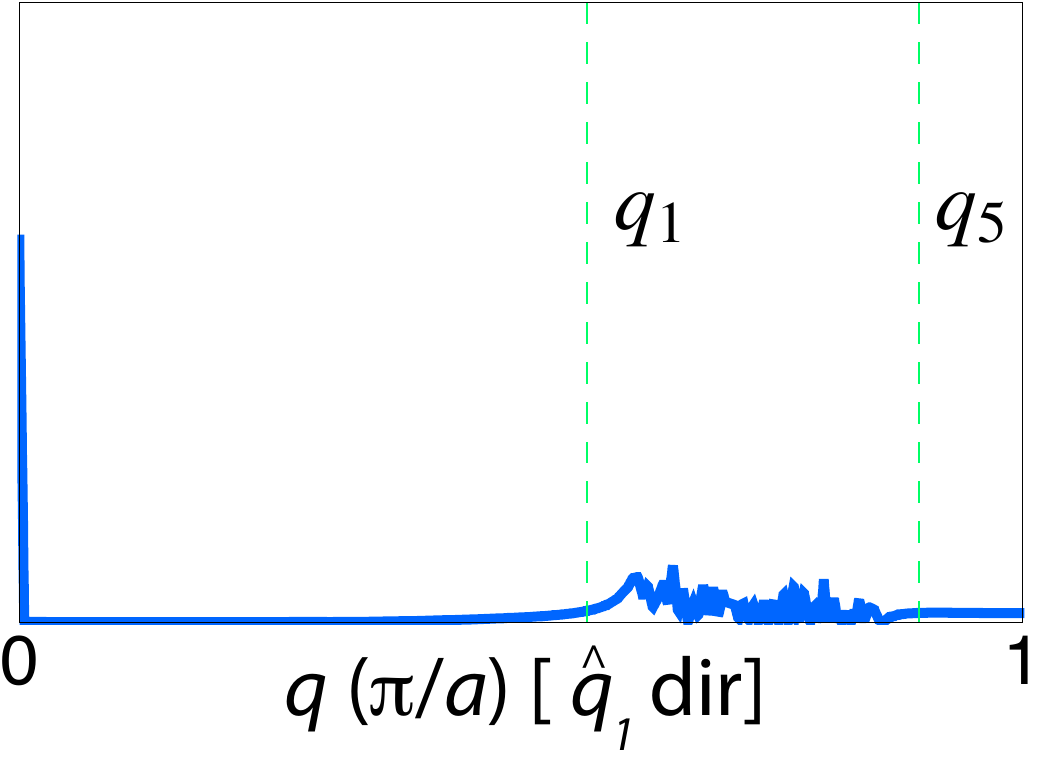}
\includegraphics[width=0.49\columnwidth]{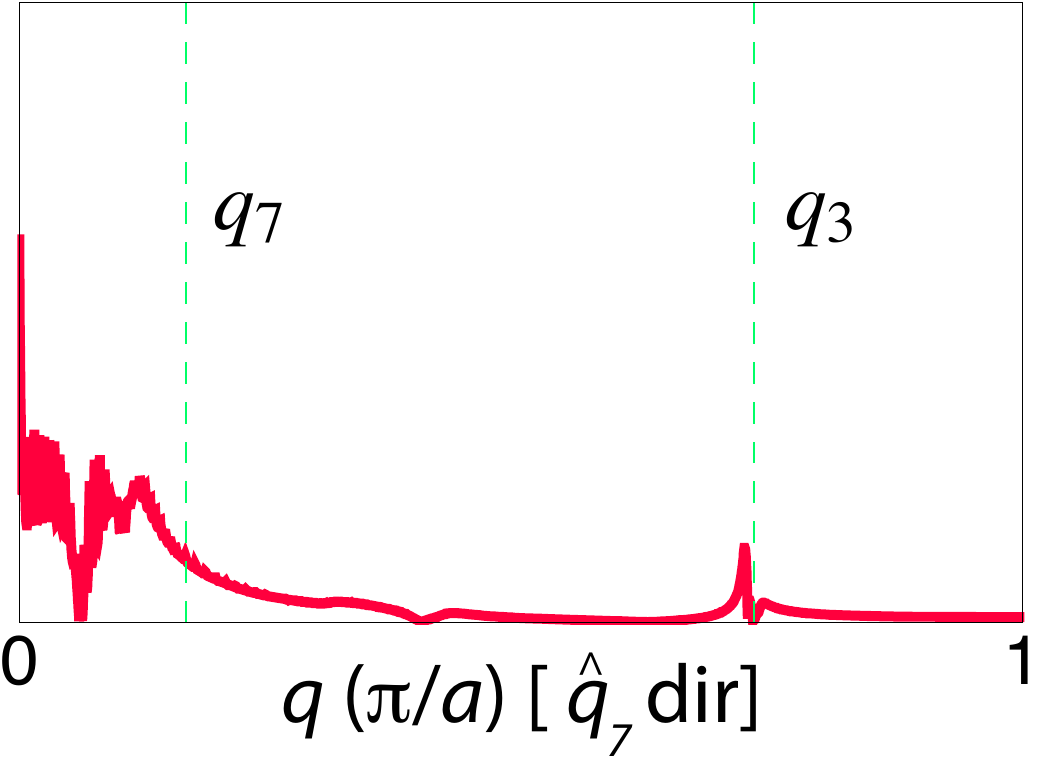}}
\subfigure[Nematic QCP, charge impurities 
]{\includegraphics[width=0.49\columnwidth]{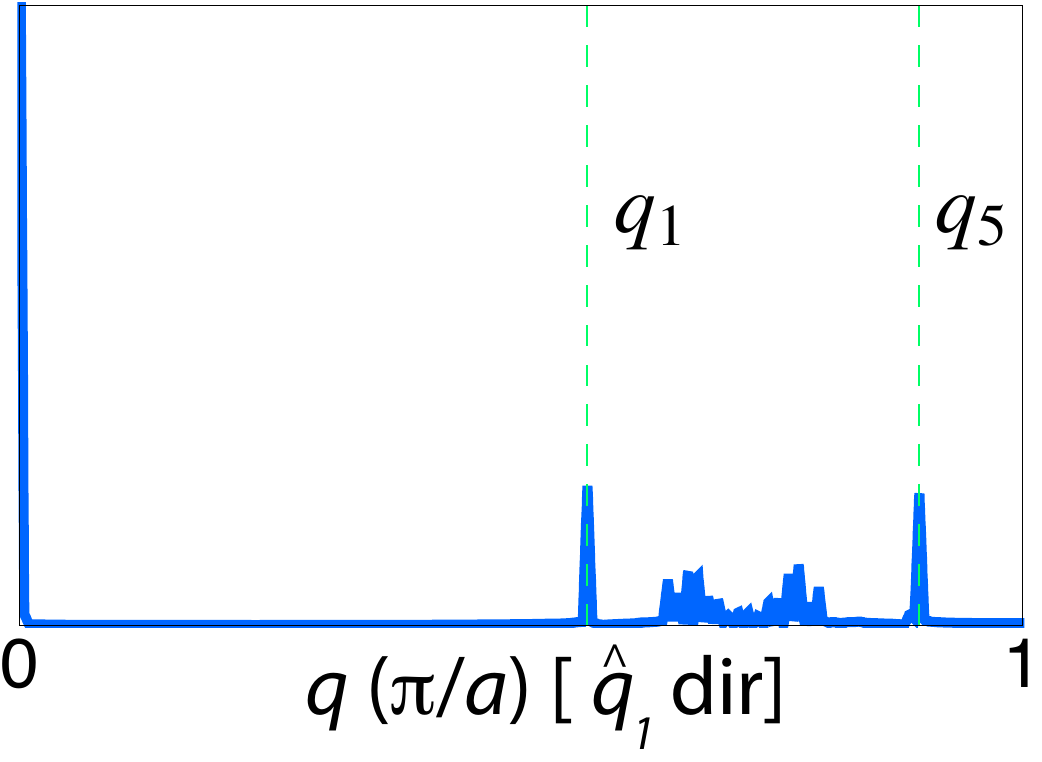}
\includegraphics[width=0.49\columnwidth]{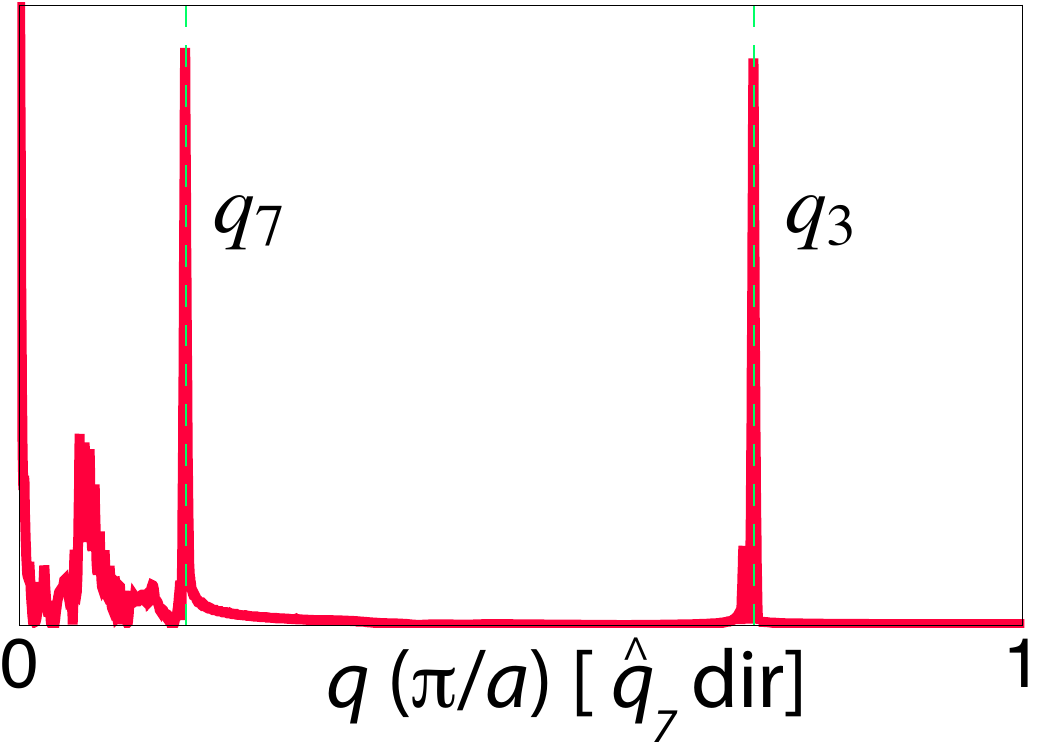}}
\subfigure[Nematic QCP, magnetic impurities
]{\includegraphics[width=0.49\columnwidth]{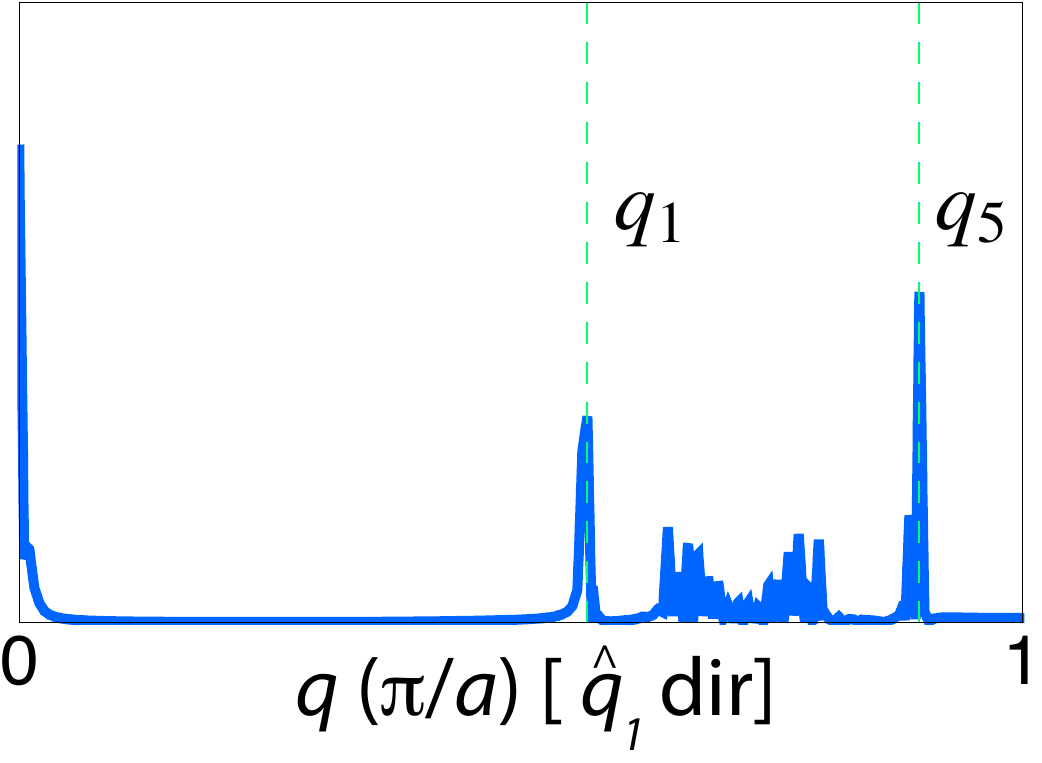}
\includegraphics[width=0.49\columnwidth]{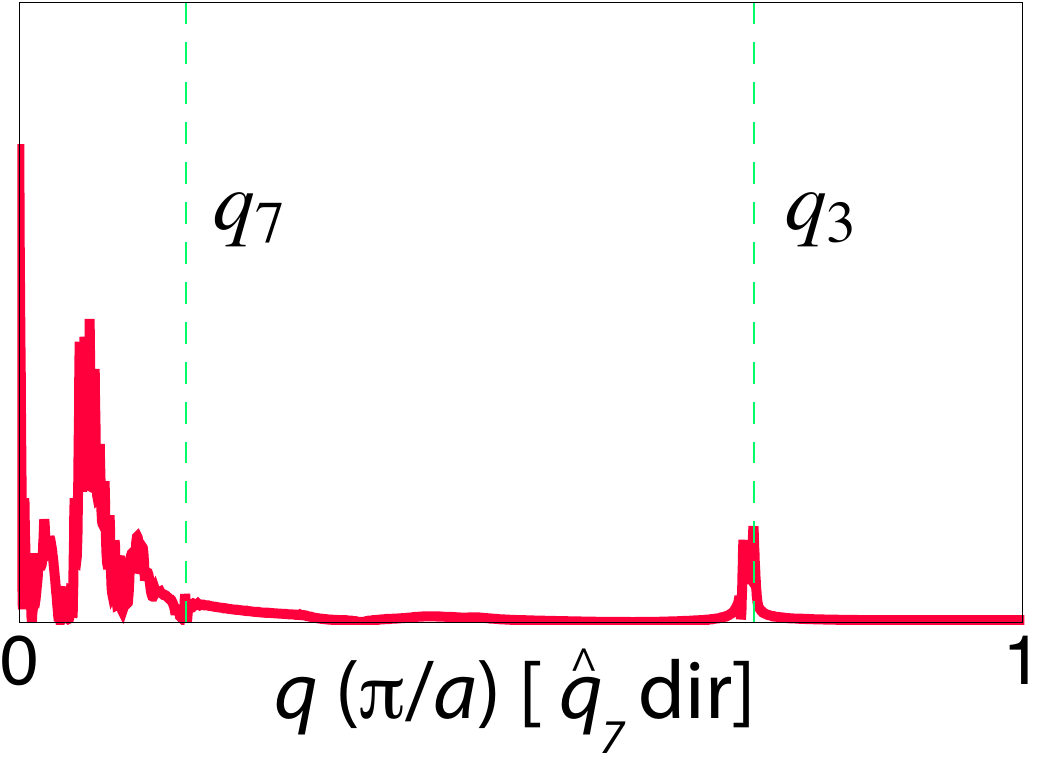}}
\caption{Normalized LDOS line cuts of the interference signal along the
$\vec q_{1}$ (or [100]) direction and along the $\vec q_{7}$ (or [110]) direction. All plots have the same y-axis scale. (a) weak features of a BCS d-wave superconductor. At the nematic QCP (b) shows scalar impurities inducing constructive interference at $\vec q_1$ and $\vec q_5$, and destructive interference at $\vec q_3$ and $\vec q_7$ while (c) shows the opposite behavior for the case of magnetic impurities.  QPI due to a pairing impurity $V_\Delta\hat{\sigma}_1$  has a qualitatively similar pattern to (c) (not presented).  }\label{fig:linecuts}
\end{figure}

The most dramatic change that nematic quantum critical fluctuations introduces in the QPI intensity, is the octet peak structure. Comparing the QPI intensity plot of Fig. \ref{fig:qpi}(b) for nematic quantum critical qp's in the presence of charge impurities to that of Fig. \ref{fig:qpi}(a) for free Bogoliubov qp's , nematic quantum critical qp's allow for unambiguous identification of all $\vec{q}_i$ vectors. (Note that the intensity is higher for sign reversing scattering vectors $\vec{q}_2$, $\vec{q}_3$, $\vec{q}_6$, and $\vec{q}_7$ when the QPI is due to charge impurities. This is consistent with the trend observed  in superconducting {Bi$_2$Sr$_2$CaCu$_2$O$_{8+\delta}$}\cite{Kohsaka:2008fj,Alldredge:2008uq,McElroy:2003mz,Kohsaka:2007fk}.)
 Such contrast between free qp's and nematic quantum critical qp's are more quantitatively displayed in line cuts in Fig.\ref{fig:linecuts} along $\vec{q}_1$ direction and $\vec{q}_7$ direction.  

The fact that nematic critical fluctuations sharpen the interference pattern and reveal positions of $\vec{q}_i$ vectors is rather striking.  One generally expects
critical fluctuations to blur the already
soft features in the interference pattern of the free Bogoliubov qp's in Fig.
\ref{fig:qpi} (a). However, the effect of the critical
nematic collective mode is quite the opposite\cite{Kim:2008gf}: it sharpens the interference pattern.
The nematic critical mode
renormalizes the dispersion of the qp's at the tips down,
reducing the gap slope $v_{\Delta}$ and enhancing the velocity
anisotropy\cite{Kim:2008gf,Huh:2008dq} 
(Intriguingly, systematic flattening of the gap slope with underdoping recently observed in 
Bi$_2$Sr$_2$CaCu$_2$O$_{8+\delta}$\cite{Tanaka:2006rg,Le-Tacon:2006le,Kohsaka:2008fj} could therefore be related to glassy nematicity).
This downward renormalization of $v_{\Delta}$ frees the qp's at the tips from further collisions with the nematic mode\cite{Kim:2008gf}.
However, qp's on the rest of the equal energy contour lose coherence after a finite time $\tau_{\vec{k}}$ set by $Im\hat{\Sigma}(\vec{k},\omega)$
that is inversely proportional to their energy $\omega$ . 

An unambiguous sign of interference origin of peaks displayed in Fig.\ref{fig:qpi}(b) would be to look for signs of constructive and destructive interference. Such distinctions can be found by noting that there are two classes of scattering  $\vec{q}_i$ vectors connecting  $\vec{k}$ space points in a $d$-wave superconductor, depending on whether they connect  points with the opposite
 sign of the gap  (red in Fig.\ref{fig:NN} and Fig.\ref{fig:qpi} ) or the same sign of the gap (blue in the same figures).
 Each of these classes of $\vec{q}_i$ vectors can be 
associated with constructive or destructive interference depending on the nature of the impurity scattering center, as it has been pointed out in Refs. \cite{Wang:2003zr,Pereg-Barnea:2003ly}.   Specifically, 
since the charged impurity potential is asymmetric in Nambu space (charged impurities affect particles and holes differently),  \textit{sign reversing} $\vec{q}_i$'s are expected to yield constructive interference.
On the other hand, both magnetic impurities $V_m\mathbb{I}$ and pair scattering centers $V_\Delta\hat{\sigma}_1$ which are symmetric in Nambu space are expected to yield constructive interference for \textit{sign preserving} $\vec{q}_i$'s. 
What is new here is that nematic critical fluctuations clearly reveal the octet peaks through the accumulation of coherence at the tips in the $\vec{k}$ space equal energy contour, and hence enable sharp comparison between QPI's induced by different types of impurity scattering centers. This reasoning is clearly borne out in the line cuts Fig.\ref{fig:linecuts}.

One way of tuning the degree of such constructive and destructive interference is to introduce vorticies.  A vortex can act both as a pair field impurity due to its core and as a magnetic impurity due to screening currents. Since both the pair field impurity and magnetic impurity are symmetric in Nambu space, vortices would shift the QPI intensity towards sign-preserving vectors as shown in Fig. \ref{fig:qpi}(c) and Fig. \ref{fig:linecuts}(c). This trend is in remarkable agreement with recent magnetic field dependence studies of QPI\cite{Hanaguri}. Scattering due to thermally excited vortices should also lead to a similar trend of enhancing peaks at $\vec{q}_1$, $\vec{q}_4$ and $\vec{q}_5$ upon raising temperature.

In summary, we have shown that nematic critical fluctuations provide a natural mechanism for the accumulation of coherence that can lead to well defined peaks in the QPI map in a manner that is consistent with the existing experimental literature. This is the first case in which the octet vectors $\vec{q}_i$ were unmistakeably revealed through a straightforward calculation. One open question is how to resolve the disappearance of the dispersing QPI peaks that accompanies the emergence of nematic glass features at high energy scales\cite{Kohsaka:2008fj,Alldredge:2008uq}.  This is a subject of future study. 
However, at this point, perhaps the most tantalizing quesion one could ask would be  whether nematic quantum critical fluctuations provide the unique mechanism for the existence of dispersing QPI peaks.

\noindent{\bf Acknowledgements} We thank  J.C.\ Davis, E.\ Fradkin, J.E.\ Hoffman, S.\ Kivelson, S.\ Sachdev, J.\ Sethna, K. Shen, O. Vafek for useful discussions. We thank T.\ Hanaguri for discussions and sharing his unpublished data.


\end{document}